\documentclass[aps,prb,twocolumn,showpacs,floatfix]{revtex4}
\usepackage{amsmath,amssymb,graphicx}
\usepackage{epsfig}
\usepackage{graphicx}
\usepackage{enumerate}
\usepackage[toc,page,header]{appendix}

\newcommand{\beq}{\begin{equation}}
\newcommand{\eeq}{\end{equation}}

\newcommand{\bea}{\begin{eqnarray}}
\newcommand{\eea}{\end{eqnarray}}

\begin{document}

\title{Flux-dependent occupations and occupation difference in
geometrically symmetric and energy
degenerate double-dot Aharonov-Bohm interferometers}

\author{Salil Bedkihal$^1$, Malay Bandyopadhyay$^2$, Dvira Segal$^1$ }
\address{$^1$Chemical Physics Theory Group, Department of
Chemistry, University of Toronto, 80 Saint George St. Toronto,
Ontario, Canada M5S 3H6\\
$^2$School of Basic Sciences, Indian Institute of Technology Bhubaneswar, 751007, India}

\pacs{73.23.-b,85.65.+h,73.63.-b}

\begin{abstract}
We study the steady-state characteristics and the transient behavior
of the nonequilibrium double-dot Aharonov-Bohm interferometer using
analytical tools and numerical simulations. Our simple setup
includes noninteracting degenerate quantum dots that are coupled to
two biased metallic leads at the same strength. A magnetic flux
$\Phi$ is piercing the setup perpendicularly. As we tune the
degenerate dots energies away from the symmetric point we observe
four nontrivial magnetic flux control effects: (i) flux dependency
of the dots occupation, (ii) magnetic flux induced occupation
difference between the dots, at degeneracy, (iii) the effect of
``phase-localization" of the dots coherence holds only at the
symmetric point, while in general both real and imaginary parts of
the coherence are nonzero, and (iv) coherent evolution survives even
when the dephasing strength, introduced into our model using
B\"uttiker probe, is large and comparable to the dots energies and
the bias voltage. Moreover, not only finite dephasing strength does
not destroy the coherence features, it can provide new type of
coherent oscillations. These four phenomena take place when the dots
energies are gated, to be positioned away from the symmetric point, demonstrating
that the combination of bias voltage, magnetic flux and gating field,
can provide delicate controllability over the occupation of each of
the quantum dots, and their coherence.
\end{abstract}

\date{\today}

\maketitle

\section{Introduction}

The steady-state properties of the Aharonov-Bohm (AB) interferometer
have been intensively investigated \cite{Rev1,Imry}, with the
motivation to explore coherence effects in electron transmission
within mesoscopic and nanoscale structures \cite{Heilb1,Heilb2}.
Particularly, the role of electron-electron (e-e) interaction
effects in AB interferometry has been considered in Refs.
\cite{GefenPRB,Gurvitz,GurvitzU,Kubo,Col,Boese,OraC}, revealing,
e.g., asymmetric interference patterns \cite{GefenPRB} and the
enhancement \cite{Boese} or elimination \cite{OraK} of the Kondo physics.
Recent works further considered the possibility of
magnetic-field control in molecular transport junctions
\cite{Hod,OraM,Dhurba,MishaDhurba}.
The real-time dynamics of AB interferometers has been of recent
interest, motivated by the challenge to understand quantum dynamics,
particularly decoherence and dissipation, in open nonequilibrium
quantum systems. Studies of electron dynamics in double-dot AB
interferometers in the absence of e-e interactions have been carried
out in Refs. \cite{Tu,Ora-time,Nori}, using a non-markovian master
equation approach. The role of e-e repulsion effects on the dots
dynamics was studied numerically using a non-perturbative
method in Ref. \cite{ABSalil}.


\begin{figure}[htbp] \vspace{-10mm} \hspace{-8mm}
\rotatebox{-90}{\hbox{\epsfxsize=120mm \epsffile{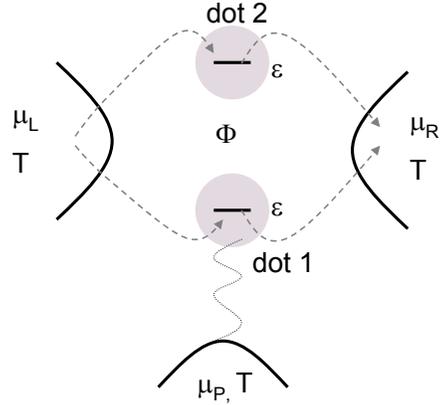}}}
\vspace{-55mm} \caption{Scheme of a double-dot AB interferometer.
The two dots are each represented by a single electronic level,
which do not directly couple. The total magnetic flux is denoted by
$\Phi$. Dot '1' may be  susceptible to dephasing effects, introduced
here through the coupling of this dot to a dephasing probe, the
terminal $P$. The role of dephasing effects is discussed in Sec. V.} \label{FigD}
\end{figure}


In this paper, we focus on a simple-minimal model, the
noninteracting double quantum dot AB interferometer, and study its
transient and steady-state properties in biased situations. For a
scheme of this model, see Fig. \ref{FigD}. This system has revealed
a wealth of intricate behavior, such as ``flux-dependent level
attraction"  \cite{Kubala}, and the ability to achieve decoherence
control when junction asymmetry is incorporated \cite{Nori}. Here,
this noninteracting system further displays other non-trivial
effects in both the transient regime and the stationary limit. While
previous studies have allowed for junction asymmetry and
nondegenerate dots \cite{Kubala,Ora-time}, we restrict ourselves to
the simplest case of energy degenerate dots and symmetric dots-lead
couplings. However, in our study the degenerate levels may be tuned
{\it away} from the symmetric point, a situation that can be reached
by applying a gating field. Using exact analytic expressions
and numerical simulations, we expose several nontrivial effects
emerging in this biased-gated AB setup: (i) First, dots' occupations
display strong flux dependency. (ii) Second, not only do the
occupations vary with flux, but the dots acquire unequal
occupations, at degeneracy. (iii) Further, we show that the effect of
``phase localization" \cite{Tu} appears only at the symmetric point,
while when the system is gated away from that point dot coherences
can be feasibly controlled by the bias voltage.

In the stationary limit we further study the role of dephasing
effects, implemented here through a dephasing probe
\cite{Buttiker1,Buttiker2,Buttiker3}, on the coherence properties of
the system. Interestingly, we find that finite and substantial
dephasing strength, at the order of the bias voltage and dots
energies, still allows for flux dependency of occupation. However,
at finite dephasing the occupation-phase dependency significantly
differs from the zero dephasing limit. In other words, finite
dephasing alter coherent oscillations, to provide new features. The
flux dependency of the occupation is fully washed out at a very
strong dephasing strength.

The structure of the paper is as follows. In Sec. II we present the
double-dot AB interferometer model. Sec. III explores the steady-state
properties of the system using the nonequilibrium Green's function
approach. We derive closed analytic expressions (at zero
temperature) for the dots' level occupation, the coherence between
the dots, and the charge current in the system. Sec. IV provides
numerical results in the transient regime, indicating on the time
scale it takes for the system to reach the stationary limit, and on
the intricate dynamics involved. Sec. V. details the role of
dephasing effects, providing analytic expressions for the dots
occupations in the steady-state limit. Sec. VI summarizes our
main results and concludes.

\section{Model}

We focus on a symmetric AB setup, with a quantum dot 
located at each arm of the interferometer. The dots are connected to
two metal leads (refereed to as baths, or reservoirs) maintained in
a biased state. For simplicity, we neglect the spin degree of
freedom and describe each quantum dot by a spinless electronic
level, see Fig. \ref{FigD}. The total Hamiltonian $H$ includes the
following terms
\beq H=H_{S}+H_{B}+H_{SB}. \label{eq:H} \eeq
Here $H_{S}$ is a subsystem (dots) Hamiltonian, $H_{B}$ includes
the two metals and $H_{SB}$ incorporates subsystem-bath interaction
terms. Specifically, we assume uncoupled dots,
\beq H_{S}=
\epsilon_{1}a_{1}^{\dagger}a_{1}+\epsilon_{2}a_{2}^{\dagger}a_{2}.
\eeq
To keep our discussion general, we allow the states to be
nondegenerate at this point. In our analytic and numerical
calculations below we have forced degeneracy. $a_{\beta}^{\dagger}$
and $a_{\beta}$ are the subsystem creation and annihilation
operators, respectively, where $\beta=1,2$ denotes dots '1' and '2'. The metals are
composed of noninteracting electrons,
\beq
H_{B}=\sum_{l}
\omega_{l}a_{l}^{\dagger}a_{l}+\sum_{r}\omega_{r}a_{r}^{\dagger}a_{r},
\eeq
where $a_{l,r}^{\dagger}$ and $a_{l,r}$ are bath creation and
annihilation operators, for the left ($l\in L$) and right ($r\in R$) leads.
The subsystem-bath interaction term is given by
\beq
H_{SB}=\sum_{\beta,l} \xi_{\beta,l} a_{\beta}^{\dagger}a_{l}
e^{i\phi_{\beta}^{L}}+\sum_{\beta,r}\zeta_{\beta,r}a_{r}^{\dagger}a_{\beta}
e^{i\phi_{\beta}^{R}}+h.c.,
\eeq
where $\xi$ is the coupling strength to left bath and similarly
$\zeta$ stands for the coupling strength to the right bath. The
notation here is general, but we later take these couplings to be
 identical, since we are interested in a dot-lead symmetric setup.
Here $\phi_{\beta}^{L}$ and $\phi_{\beta}^{R}$ are the AB phase
factors, acquired by electron waves in a magnetic field
perpendicular to the device plane. These phases are constrained to
satisfy the following relation
\beq
\phi_{1}^{L}-\phi_{2}^{L}+\phi_{1}^{R}-\phi_{2}^{R}=\phi=2\pi\Phi/\Phi_{0}.
\eeq
$\Phi$ is the magnetic flux enclosed by the ring and $\Phi_{0}=hc/e$
is the flux quantum. In what follows we adopt the gauge
$\phi_{1}^{L}-\phi_{2}^{L}=\phi_{1}^{R}-\phi_{2}^{R}=\phi/2$.

We voltage-bias the system, using the convention $\Delta
\mu\equiv\mu_L-\mu_R\geq0$, with $\mu_{L,R}$ as the chemical
potential of the metals. While we bias the system in a symmetric
manner, $\mu_L=-\mu_R$, the dots levels may be placed away from the
so called ``symmetric point" at which
$\mu_L-\epsilon_{\beta}=\epsilon_{\beta}-\mu_R$. This situation may
be achieved by applying a gate voltage to each dot.
For simplicity, we use the following conventions below, $\hbar\equiv 1$ and electron charge
$e\equiv 1$.


\section{Stationary behavior}

\subsection{Method: equations of motion}

Since the model is noninteracting, its steady-state characteristics can
be calculated exactly using the nonequilibrium Green's function
(NEGF) approach \cite{Wingreen}. This technique has been  
extensively used in the past for studying transport properties in mesoscopic systems
and molecular junctions \cite{RevGreen}. We review here the steps
involved so as to carefully contain the phase factors. The
derivation presented here follows an equation-of-motion approach
\cite{sen}. In this method, an effective quantum Langevin equation
for the subsystem is obtained by solving the Heisenberg equations of
motion (EOM) for the baths variables, then substituting them back
into the EOM for the subsystem (dots) variables. The indices
$\alpha,\beta=1,2$ identify the two dots. The resulting EOM is
\begin{eqnarray}
\frac{da_{\beta}}{dt}&=&-i\epsilon_{\beta}a_{\beta}-i\eta_{\beta}^{L}-i\eta_{\beta}^{R}
\nonumber\\
&-&i\int_{t_{0}}^{t}d\tau \sum_{\alpha,l} \xi_{\beta,l}
{g_{l}^{+}}\left(t-\tau\right)\xi_{\alpha,l}^{*}e^{i(\phi_{\beta}^{L}-\phi_{\alpha}^{L})}
a_{\alpha}(\tau)
\nonumber\\
&-&i\int_{t_{0}}^{t}d\tau\sum_{\alpha,r} \zeta_{\beta,r}^{*}
{g_{r}^{+}}\left(t-\tau\right)
\zeta_{\alpha,r}e^{i(\phi_{\alpha}^{R}-\phi_{\beta}^{R})}
a_{\alpha}(\tau).
\nonumber\\
\label{eq:EOMa}
\end{eqnarray}
The (isolated) reservoirs Green's functions are given by
\begin{eqnarray}
 g^{+}_l(t)=-ie^{-i\omega_lt}\theta(t),\,\,\,\,\,
 g^{+}_r(t)=-ie^{-i\omega_rt}\theta(t).
\end{eqnarray}
The terms $\eta_{\beta}^{L}$ and $\eta_{\beta}^{R}$ are referred to as
{\it noise}, induced on the subsystem from the left and right
reservoirs, respectively. Their explicit form is
\begin{eqnarray}
\eta_{\beta}^{L}&=& i\sum_{l}\xi_{\beta,l} g_{l}^{+}\left(t-t_{0}\right)a_{l}(t_{0})e^{i\phi_{\beta}^{L}}
\nonumber\\
\eta_{\beta}^{R}&=&i\sum_{r}\zeta_{\beta,r}^{*}g_{r}^{+}\left(t-t_{0}\right)a_{r}(t_{0})e^{-i\phi_{\beta}^{R}}.
\end{eqnarray}
As an initial condition, we take a factorized state,
$\rho(t_0)=\rho_L\otimes \rho_R\otimes \rho_S(t_0)$, with empty dots and the
reservoirs prepared in a grand canonical state,
$\rho_{\nu}=\frac{e^{-(H_{\nu}-\mu_{\nu} N)/T_{\nu}}}{{\rm
Tr}[\rho_{\nu}]}$, $T_{\nu}$ is the temperature of the $\nu=L,R$
fermi sea and $\mu_{\nu}$ stands for its chemical potential. The
reduced density matrix $\rho_S$ denotes the state of the subsystem.
Using this initial condition, noise correlations satisfy
\begin{eqnarray}
&&\langle \eta_{\beta}^{\dagger{L}}(t)
\eta_{\beta'}^{L}(\tau)\rangle=
\sum_{l} \xi_{\beta,l}^{*}
e^{i\omega_{l}(t-\tau)}
\xi_{\beta',l}e^{-i(\phi_{\beta}^{L}-\phi_{\beta'}^{L})}
f_L(\omega_{l})
\nonumber\\
&&\langle \eta_{\beta}^{\dagger{R}}(t)
\eta_{\beta'}^{R}(\tau)\rangle
= \sum_{r} \zeta_{\beta,r}
e^{i\omega_{r}(t-\tau)}
\zeta_{\beta',r}^*e^{i(\phi_{\beta}^{R}-\phi_{\beta'}^{R})}f_R(\omega_{r}),
\nonumber
\end{eqnarray}
%
with the Fermi function
$f_{\nu}(\omega)=[e^{(\omega-\mu_{\nu})/T_{\nu}}+1]^{-1}$ and
expectation values evaluated in the  Heisenberg representation,
$\langle A(t) \rangle={\rm Tr}[\rho(t_0)A(t)]$.
Steady-state properties are reached by taking the limits
$t_0\rightarrow -\infty$ and $t\rightarrow \infty$. We now Fourier
transform Eq. (\ref{eq:EOMa}) using the convolution theorem with the
convention $\tilde a_{\beta}(\omega)=\int_{-\infty}^{\infty} dt
a_{\beta}(t)e^{i\omega t}$,  $\tilde \eta_{\beta}(\omega)=
\int_{-\infty}^{\infty} dt \eta_{\beta}(t)e^{i\omega t}$. The
result, organized in a matrix form, is
\begin{equation}
\tilde a_{\beta}(\omega)=\sum_{\alpha} G^{+}_{\beta,\alpha}\lbrack
\tilde{\eta}^L_{\alpha}(\omega) +
\tilde{\eta}^R_{\alpha}(\omega)\rbrack,
\label{eq:sol}
\end{equation}
with the Green's function
\begin{equation}
G^{+}_{\beta,\alpha}(\omega)=
\frac{1}{(\omega-\epsilon_{\beta})\delta_{\alpha,\beta}-\Sigma_{\beta,\alpha}^{L,+}(\omega)-\Sigma_{\beta,\alpha}^{R,+}(\omega)}.
\end{equation}
The self energies contain the phase factors,
\begin{eqnarray}
\Sigma_{\beta,\alpha}^{L,+}(\omega)&=&\sum_{l}\xi_{\beta,l}g^{+}_{l}(\omega)
\xi_{\alpha,l}^{*} e^{i(\phi_{\beta}^{L}-\phi_{\alpha}^{L})},
\,\,\,\,\,
\nonumber\\
\Sigma_{\beta,\alpha}^{R,+}(\omega)&=&\sum_{r}\zeta_{\beta,r}^*g^{+}_{r}(\omega)
\zeta_{\alpha,r} e^{i(\phi_{\alpha}^{R}-\phi_{\beta}^{R})}.
\end{eqnarray}
%
We also define the conjugated-transposed matrix,
$G^{-}=({G^{+}})^{\dagger}$, to be used below.
The real part of the self energy is a principal value integral,
assumed here to vanish. This assumption holds when the metals density of states
is energy independent and the bandwidth is large.
We then define the hybridization matrix from the relation $\Sigma^+=-i\Gamma/2$,
\begin{eqnarray}
\Gamma^{L}_{\beta,\beta'}(\omega)=2\pi
e^{i(\phi_{\beta}^{L}-\phi_{\beta'}^{L})}\sum_l \xi_{\beta,l}
\delta(\omega-\omega_l) \xi_{\beta',l}^{*}.
\label{eq:Gamma}
\end{eqnarray}
%
Similar expressions hold for the $R$ side. Using the steady-state
solution (\ref{eq:sol}), we can write down an expression for the reduced
density matrix.  Back transformed to the time domain it takes the form
\bea
\langle a_{\alpha}^{\dagger}a_{\beta}\rangle &\equiv& \rho_{\alpha,\beta} =
\frac{1}{2\pi}\int_{-\infty}^{\infty}\Big[\left(G^{+}\Gamma^{L}G^{-}\right)_{\alpha,\beta}
f_L(\omega)
\nonumber\\
&+&\left(G^{+}\Gamma^{R}G^{-}\right)_{\alpha\beta}f_R(\omega)\Big]d\omega.
\label{eq:rho}
\eea
The time variable has been suppressed since the result is only valid
in the steady-state limit. In what follows we take $\xi_{\beta,l}$
and $\zeta_{\beta,r}$ as real constants, independent of the level
index and the reservoir state, resulting in 
\bea
\Gamma^L_{\beta,\beta'}=\gamma_Le^{i(\phi_{\beta}^L-\phi_{\beta'}^L)}, \,\,\,\
\Gamma^R_{\beta,\beta'}=\gamma_Re^{-i(\phi_{\beta}^R-\phi_{\beta'}^R)},
\eea
where the coefficient $\gamma_{\nu}$, defined through this relation and Eq.
(\ref{eq:Gamma}), is taken as a constant, energy independent. Using
these definitions, the matrix $G^{+}$ takes the form
\bea G^{+}=\left[  \begin{array}{cc}
 \omega-\epsilon_{1}+\frac{i(\gamma_L+\gamma_R)}{2} & \frac{i\gamma_L}{2}e^{i\phi/2}
+\frac{i\gamma_R}{2}e^{-i\phi/2}\\
\frac{i\gamma_L}{2}e^{-i\phi/2}+\frac{i\gamma_R}{2}e^{i\phi/2}
&  \omega-\epsilon_{2}+\frac{i(\gamma_L+\gamma_R)}{2}\\
 \end{array}\right]^{-1}
\label{eq:G1}
\eea
and the hybridization matrices are given by
\begin{equation} \Gamma^{L}=\gamma_L\left[ \begin{array}{cc}
              1 & e^{i\phi/2}\\
              e^{-i\phi/2} &1\\
\end{array}\right],\,\,\,\,\,\,\,\,\,\,\,
\Gamma^{R}=\gamma_R \left[ \begin{array}{cc}
              1 & e^{-i\phi/2}\\
              e^{i\phi/2} &1\\
\end{array}\right]
\end{equation}
We can now calculate, numerically or analytically, the behavior of
the reduced density matrix under different conditions
\cite{Ora-time}. Since we are only concerned here with {\it
symmetric} dot-lead couplings, we take $\gamma_L=\gamma_R=\gamma/2$.
Furthermore, we impose energy degeneracy, $\epsilon_1=
\epsilon_2=\epsilon$. This choice simplifies
the relevant matrices to
\bea
G^{+}=\left[
\begin{array}{cc}
\omega-\epsilon+\frac{i\gamma}{2} & \frac{i\gamma}{2}\cos\frac{\phi}{2} \\
\frac{i\gamma}{2}\cos\frac{\phi}{2}
&  \omega-\epsilon+\frac{i\gamma} {2}\\
 \end{array}
\right]^{-1},\,\,\,\,\,\,\,
\nonumber\\
\Gamma^{L}=\frac{\gamma}{2} \left[
\begin{array}{cc}
              1 & e^{i\phi/2}\\
              e^{-i\phi/2} &1\\
 \end{array}
\right],\,\,\,\,
\Gamma^{R}=\frac{\gamma}{2} \left[
\begin{array}{cc}
          1 & e^{-i\phi/2}\\
              e^{i\phi/2} &1\\
 \end{array}
\right]
\label{eq:Gs}
\eea
We present closed analytic expressions for the diagonal and
off-diagonal elements of the reduced density matrix in Sec. III.B.
Complementing numerical data for the real-time dynamics are included
in Sec. IV. This discussion is generalized in Sec. V, to include a dephasing probe.


\subsection{Observables}

\subsubsection{Dots occupation}

We expose here two effects that persist away from the ``symmetric
point", defined as $\mu_L-\epsilon=\epsilon-\mu_R$:  The dots
occupations significantly vary with flux, and moreover, the degenerate
dots acquire different occupations. After presenting general
expressions away from the symmetric point, we consider other
relevant cases: the finite-bias limit at the symmetric point, the
limit of infinite bias (which effectively reduces to the symmetric
point), and the case of $\phi=2\pi n$, $n=0,1,2$...

\begin{figure}
\hspace{2mm} 
{\hbox{\epsfxsize=75mm \epsffile{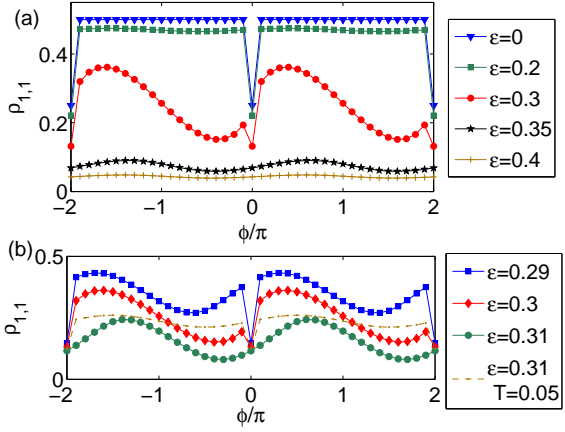}}} \caption{(a) Flux
dependency of occupation for dot '1' using $\epsilon=0$ (triangle)
$\epsilon=0.2$ ($\square$) $\epsilon=0.3$ ($\circ$), $\epsilon=0.35$
($\star$) and $\epsilon=0.4$ ($+$). Panel (b) displays results when
$\epsilon$ is tuned to the bias window edge, $\epsilon\sim \mu_L$,
$\epsilon=0.29$ ($\square$), $\epsilon=0.3$ (diagonal),
$\epsilon=0.31$ ($\circ$), and $\epsilon=0.31$, $T=0.05$
(dashed-dotted line). In all cases $\mu_L=-\mu_R=0.3$,
$\gamma=0.05$, and $T=0$, unless otherwise stated.} \label{Figpop0}
\end{figure}

Analytic results are obtained from Eqs.
(\ref{eq:rho}) and (\ref{eq:Gs}). Organizing these expressions, we
find that the occupation of dot '1', $\rho_{1,1}\equiv\langle
a_1^{\dagger} a_1\rangle$, is given by two integrals,
\begin{eqnarray}
\rho_{1,1}=
\frac{\gamma}{4\pi}
\int_{-\infty}^{\infty}f_L(\omega)d\omega
\frac{ (\omega-\epsilon)^2+\omega_0^2-
2\omega_0(\omega-\epsilon)\cos\frac{\phi}{2}}
{\left[(\omega-\epsilon)^2-\omega_0^2\right]^2+ [\gamma(\omega-\epsilon)]^2 }
\nonumber \\
+ \frac{\gamma}{4\pi}
\int_{-\infty}^{\infty}f_R(\omega)d\omega
\frac{(\omega-\epsilon)^2+\omega_0^2 + 2\omega_0(\omega-\epsilon)\cos\frac{\phi}{2}}
{\left[(\omega-\epsilon)^2-
\omega_0^2\right]^2+ [\gamma(\omega-\epsilon)]^2}, \,\,\,\,\
\label{eq:p1E}
\end{eqnarray}
where we have introduced the short notation
\bea \omega_0\equiv \frac{\gamma}{2}\sin\frac{\phi}{2}. \eea
Similarly, the occupation of level '2', $\rho_{2,2}\equiv \langle
a_2^{\dagger}a_2\rangle $, is given by
\begin{eqnarray}
\rho_{2,2}=
\frac{\gamma}{4\pi}
\int_{-\infty}^{\infty}f_L(\omega)d\omega
\frac{ (\omega-\epsilon)^2+\omega_0^2+
2\omega_0(\omega-\epsilon)\cos\frac{\phi}{2}}
{\left[(\omega-\epsilon)^2-\omega_0^2\right]^2+ [\gamma(\omega-\epsilon)]^2 }
\nonumber \\
+ \frac{\gamma}{4\pi}
\int_{-\infty}^{\infty}f_R(\omega)d\omega
\frac{(\omega-\epsilon)^2+\omega_0^2 - 2\omega_0(\omega-\epsilon)\cos\frac{\phi}{2}}
{\left[(\omega-\epsilon)^2-
\omega_0^2\right]^2+ [\gamma(\omega-\epsilon)]^2}. \,\,\,\,\
\label{eq:p2E}
\end{eqnarray}
In what follows we consider the zero temperature limit.
The Fermi functions take then the shape of step functions and
the upper limits of the integrals are replaced by the corresponding chemical potentials.
We now study the contribution of the odd term in the integrand. This
term is responsible for the development of occupation difference
between the dots,
\bea
&&\frac{\gamma}{4\pi}\int_{\mu_R}^{\mu_L}d\omega
\frac{2\omega_0(\omega-\epsilon)\cos\frac{\phi}{2}}
{\left[(\omega-\epsilon)^2-\omega_0^2\right]^2+ [\gamma(\omega-\epsilon)]^2 }
\nonumber\\
&&=
\frac{\sin\frac{\phi}{2}}{8\pi}\ln\left[\frac{F_+(\phi)}{F_-(\phi)}\right],
\label{eq:F1}
\eea
where the explicit form of the factors $F_{\pm}$ is
%
%
%
\bea
F_{\pm}(\phi)&=& \frac{\gamma^4}{8}\sin^4 \frac{\phi}{2}+2(\mu_L-\epsilon)^2(\mu_R-\epsilon)^2
\nonumber\\
&+&\frac{\gamma^2}{2}\left(\cos\frac{\phi}{2}\pm 1\right)^2 (\mu_L-\epsilon)^2
\nonumber\\
&+&\frac{\gamma^2}{2}\left(\cos\frac{\phi}{2}\mp 1\right)^2 (\mu_R-\epsilon)^2.
\label{eq:F3}
\eea
For details, see Appendix A.
Since it is a sum of real quadratic terms, $F_{\pm}\geq0$.
Inspecting Eq. (\ref{eq:F1}), we note that it vanishes in four
different cases: (i) at zero bias, when $\mu_L=\mu_R=0$, (ii) at
infinite bias, $\mu_L\rightarrow \infty$ and $\mu_R\rightarrow
-\infty$, (iii) at the symmetric point when
$\mu_L-\epsilon=\epsilon-\mu_R$, particularly for $\epsilon=0$ and
$\mu_L=-\mu_R$, or when (iv) $\phi=n\pi$, $n=0,1,2..$ (leading to
$F_+=F_-$). Combining Eq. (\ref{eq:F1}) with the integration of even
terms in Eq. (\ref{eq:p1E}), at zero temperature, we resolve the occupations
\begin{eqnarray}
\rho_{1,1/2,2}&=&\frac{1}{4\pi}\Big\lbrack 2\pi
+ \tan ^{-1}\Big(\frac{\mu_L-\epsilon}{\gamma_{-}}\Big)
+\tan ^{-1}\Big(\frac{\mu_L-\epsilon}{\gamma_{+}}\Big)
\nonumber\\
&+& \tan ^{-1}\Big(\frac{\mu_R-\epsilon}{\gamma_{-}}\Big)
+\tan ^{-1}\Big(\frac{\mu_R-\epsilon}{\gamma_{+}}\Big) \Big\rbrack
\nonumber \\
&\pm&\frac{\sin\frac{\phi}{2}}{8\pi}\ln\left[\frac{F_-(\phi)}{F_+(\phi)}\right].
\label{eq:p1}
\end{eqnarray}
The positive sign corresponds to $\rho_{1,1}$, the negative sign
provides $\rho_{2,2}$. We have also introduced the short notation
$\gamma_{\pm}\equiv\frac{\gamma}{2}(1\pm\cos\frac{\phi}{2})$.
Equation (\ref{eq:p1}) predicts flux dependency of electron
occupation {\it at degeneracy}, using symmetric hybridization
constants, once the dots are tuned {\it away} from the symmetric
point. Fig. \ref{Figpop0} displays this behavior, and we find that
as the dots energies get closer to the bias edge,
$\epsilon\sim\mu_L$, the population strongly varies with $\epsilon$
(panel b). It is also interesting to note that the abrupt jump at
$\phi=2\pi n$ (discussed below) disappears once the levels reside at
or above the bias window, for $\epsilon\geq\mu_L$. This feature
results from the strict zero temperature limit assumed in the
analytic calculations. At finite $T$ the jump at $\phi=2\pi n$
survives even for $\epsilon>\mu_L$. However, when the temperature is
at the order of the hybridization strength, $T\sim \gamma$, the
modulation of the population with phase is washed out. The following
parameters are used here and below: flat wide bands, dots energies
at the order of $\epsilon=0-0.4$, hybridization strength
$\gamma=0.05-0.5$, and a zero temperature, unless otherwise
specified. The bias voltage is set symmetrically around the
equilibrium Fermi energy, $\mu_L=-\mu_R$, $\Delta
\mu\equiv\mu_L-\mu_R$.


\begin{figure}
\hspace{2mm} 
{\hbox{\epsfxsize=85mm \epsffile{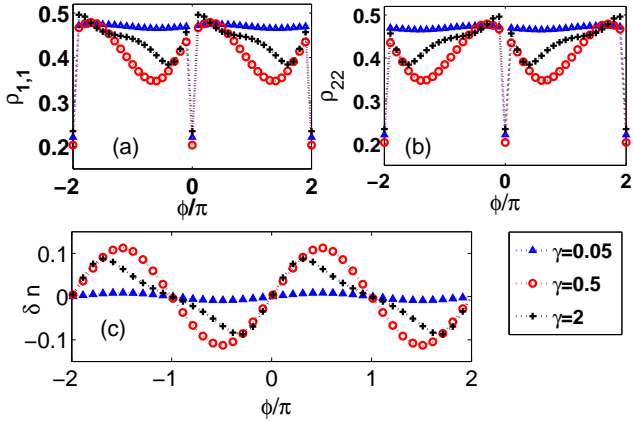}}} \caption{ (a)-(b) Dots
occupations as a function of magnetic phase $\phi$ for $\Delta
\mu=0.6$, $\epsilon=0.2$, $T=0$. (c) Occupation difference, $\delta
n=\rho_{1,1}-\rho_{2,2}$. At weak coupling, $\gamma=0.05$
($\triangle$), the dots occupations are almost identical. When the
hybridization is made stronger, $\gamma=0.5$ ($\circ$), comparable
to the levels displacement from the symmetric point, $\rho_{1,1}$
clearly deviates from $\rho_{2,2}$. At very strong coupling,
$\gamma=2$ ($+$), the occupation difference reduces and asymmetries
develop. For clarity, results are shown for $\phi/\pi$ between
(-2,2).} \label{Figpop}
\end{figure}


\begin{figure}
\hspace{2mm} 
{\hbox{\epsfxsize=75mm \epsffile{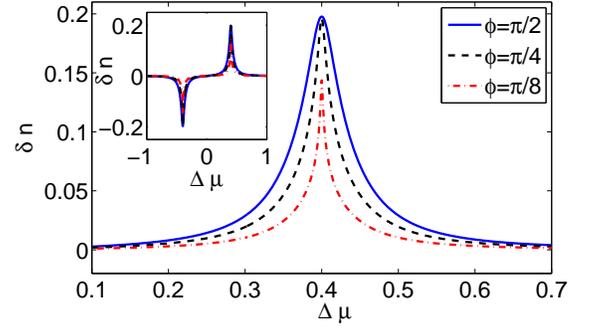}}} \caption{Occupation
difference as a function of bias voltage $\Delta \mu$, for different magnetic
flux values, $\phi=\pi/2$ (full line), $\phi=\pi/4$ (dashed line),
$\phi=\pi/8$ (dashed-dotted line). Other parameters are
$\epsilon=0.2$ and $\gamma=0.05$, $T=0$. The inset presents data
for backward and forward biases; the main plot zooms on the
positive bias regime.} \label{Figp1}
\end{figure}

\begin{figure}
\hspace{2mm} 
{\hbox{\epsfxsize=65mm \epsffile{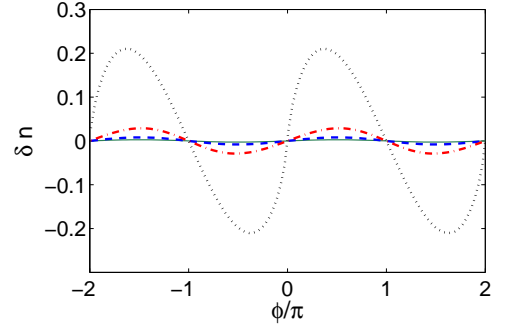}}} \caption{Occupation
difference as a function of magnetic flux for different bias values,
$\Delta \mu=0.1$ (full line), $\Delta \mu=0.2$ (dashed line),
$\Delta \mu=0.3$ (dashed-dotted line) and $\Delta \mu=0.4$ (dotted
line). Other parameters are $\epsilon=0.2$, $\gamma=0.05$ and $T=0$.}
\label{Figp2}
\end{figure}


We now discuss in more details the behavior of the occupation in
some special cases. First, we consider the symmetric point at finite
bias and $\phi\neq 2\pi n$, $n=0,1,2...$. In this case Eq.
(\ref{eq:p1}) precisely reduces to
\bea
\rho_{\alpha,\alpha}(\mu_L-\epsilon=\epsilon-\mu_R)=\frac{1}{2}.
\label{eq:p1II}
\eea
This result holds in the infinite bias limit,
$\mu_L\rightarrow \infty$ and $\mu_R\rightarrow -\infty$,
irrespective of the (finite) value of $\epsilon$. Next, the special
case $\phi=2\pi n$ should be separately evaluated. At these points we have
$\omega_0=0$ and Eq. (\ref{eq:p1E}) provides the simple form at zero temperature
\bea
&&\rho_{\alpha,\alpha}(\phi=2  \pi n )
\nonumber\\
&&=
\frac{\gamma}{4\pi}\int_{-\infty}^{\mu_L-\epsilon}  \frac{dx}{x^2+\gamma^2}
+ \frac{\gamma}{4\pi}\int_{\epsilon-\mu_R}^{\infty}  \frac{dx}{x^2+\gamma^2}
\nonumber\\
&&= \frac{1}{4\pi} \left[ \tan^{-1}\left(\frac{\mu_L-\epsilon}{\gamma} \right)
 + \tan^{-1}\left(\frac{\mu_R-\epsilon}{\gamma} \right) \right]
+\frac{1}{4}.
\label{eq:p1III}
\nonumber\\
\eea
These points are reflected by abrupt jumps in the occupations-flux
behavior. Specifically, at the symmetric point there is a sharp
reduction of occupation number from $1/2$ [Eq. (\ref{eq:p1II})] to
$1/4$ [Eq. (\ref{eq:p1III})], as observed earlier in Ref. \cite{Tu}.
Fig. \ref{Figpop0} shows that at strictly zero temperature this jump
disappears once the dots energies are placed at or above the bias
edge, $\epsilon\geq \mu_L$. Thus, the appearance of the jump is indicative of
the fact that
electrons cross the junction resonantly. 
If only tunneling processes contribute 
(once the dots' energies are placed above the bias window and the 
temperature is very low), the population continuously vary with flux.

The total electronic occupation of the dots, at steady-state, generalizes the standard
symmetric case attained in Ref. \cite{Ora-time},
\bea
\rho_{1,1}+\rho_{2,2}=
\frac{\gamma}{2\pi}\int_{-\infty}^{\infty} d\omega
\frac{[(\omega-\epsilon)^2+\omega_0^2] \left[f_L(\omega)+f_R(\omega) \right]
}
{\left[(\omega-\epsilon)^2-\omega_0^2\right]^2 + [\gamma(\omega-\epsilon)]^2}.
\nonumber\\
\eea
We now highlight one of the main results of the paper, the onset of
occupation difference in this degenerate ($\epsilon_1=\epsilon_2$)
and spatially symmetric ($\gamma_L=\gamma_R$) setup. Using Eq.
(\ref{eq:p1}), we find that
\bea \delta n\equiv
\rho_{1,1}-\rho_{2,2}=\frac{\sin\frac{\phi}{2}}{4\pi}\ln\left[
\frac{F_-(\phi)}{F_+(\phi)}\right].
\eea
%
As we mentioned above, this quantity is nonzero when the following
(sufficient) conditions are simultaneously satisfied: (i) the bias
voltage is finite, neither zero nor infinite, (ii) the dots are
positioned away from the symmetric point,
$\epsilon\neq(\mu_L+\mu_R)/2$, and (ii) the phase $\phi$ is not a
multiple of $\pi$, $\phi\neq n\pi$, $n=0,1,2...$. To rephrase this
observation, 
the occupation difference can be controlled by manipulating the
subsystem-metal hybridization energy $\gamma$, by changing the bias
voltage, by applying a gate voltage for tuning 
the dots energies, and by modulating the phase $\phi$ through the magnetic flux. 
The role of these control knobs are illustrated in Figures \ref{Figpop},
\ref{Figp1} and \ref{Figp2}.

In Fig. \ref{Figpop} we display the levels occupation in the
resonant regime, $\mu_R<\epsilon<\mu_L$ while varying $\gamma$. At
weak coupling $\delta n$ is insignificant. However, the occupation
difference becomes large when co-tunneling effects contribute. More
notably, Fig. \ref{Figp1} illustrates the strong controllability of
$\delta n$ with applied voltage. We find that the occupation
difference is maximized at the edge of the resonant transmission
window, when $\mu_L-\epsilon=0$ (or equivalently, when $\Delta
\mu=2\epsilon$). The magnetic phase affects the width and height of
the peak, but not the absolute position which is only
determined by the offset of $\epsilon$ from the center of the bias
window. In Fig. \ref{Figp2} we further show the flux dependency of
$\delta n$, which is particularly significant when $\Delta
\mu=2\epsilon$.

\begin{figure}
\hspace{2mm} 
{\hbox{\epsfxsize=70mm \epsffile{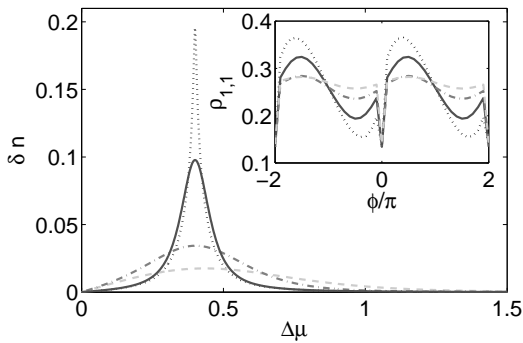}}} \caption{Finite
temperature effect.
Main plot: Occupation difference as a
function of bias voltage for $\phi=\pi/4$. Inset: 
occupation of dot '1' as a function of magnetic phase for $\Delta \mu=0.4$.
In both panels
$T=0$ (dotted line), $T=0.01$ (full line), $T=0.05$ (dashed-dotted
line) and $T=0.1$ (dashed line). Dots parameters are $\epsilon=0.2$
and $\gamma=0.05$.} \label{FigTT}
\end{figure}

The effect of finite temperature on the occupation-flux dependence,
and on the development of occupation difference, is displayed in Fig.
\ref{FigTT}. We find that the effects largely survive at finite $T$,
as long as $T<\gamma$. These results were calculated numerically,
based on Eqs. (\ref{eq:p1E}) and (\ref{eq:p2E}).


\subsubsection{Coherence}

It was recently argued that the decoherence behavior in our generic
setup, including two noninteracting (uncoupled) quantum dots
interferometer, can be suppressed when the device geometry is made
asymmetric and nondegenerate, using $\epsilon_1\neq \epsilon_2$ and
$\gamma_L\neq \gamma_R$ \cite{Nori}. The requirement for asymmetry
in this work arises from the observation of the ``phase
localization"  effect, which hinders phase manipulation in the
system at the symmetric point. The term ``phase localization" refers
to the fact that if we define
$\rho_{1,2}(t)=|\rho_{1,2}(t)|e^{i\varphi(t)}$, the relative phase
$\varphi$ localizes to the values $-\pi/2$ or $\pi/2$ in the long
time limit when $\phi\neq 2\pi n$, $n$ is an integer \cite{Tu}.
Based on numerical simulations, we have pointed out in Ref.
\cite{ABSalil} that phase localization occurs {\it only} at the
symmetric point, while at other values of $\epsilon$ the real part
of $\rho_{1,2}$ is finite and nonzero in the asymptotic limit for
any phase besides $2\pi n$ \cite{ABSalil}. This observation is
established here analytically in the steady-state limit, implying
that decoherence could be suppressed in degenerate-symmetric systems
by gating the dots, shifting their energies relative to the bias
window.

We derive a closed expression for the off-diagonal system element
$\rho_{1,2}\equiv \langle a_1^{\dagger}a_2\rangle$ by studying Eq.
(\ref{eq:rho}),
\begin{widetext}
\begin{eqnarray}
\rho_{12}&=&\frac{\gamma}{4\pi}
\int_{-\infty}^{\infty} f_L(\omega)d\omega
\left\{\frac{\cos{\frac{\phi}{2}} \left[(\omega-\epsilon)^2-\omega_{0}^{2}\right]
+ i\sin{\frac{\phi}{2}}\left[(\omega-\epsilon)^2+\omega_{0}^{2}\right]}
{\left[(\omega-\epsilon)^2-\omega_{0}^2\right]^2+\left[\gamma(\omega-\epsilon)\right]^2}\right\}
\nonumber\\
&&+ \frac{\gamma}{4\pi}
\int_{-\infty}^{\infty} f_R(\omega)d\omega
\left\{\frac{\cos{\frac{\phi}{2}} \left[(\omega-\epsilon)^2-\omega_{0}^{2}\right]
-i\sin{\frac{\phi}{2}}\left[(\omega-\epsilon)^2+\omega_{0}^{2}\right]}
{\left[(\omega-\epsilon)^2-\omega_{0}^2\right]^2+\left[\gamma(\omega-\epsilon)\right]^2}\right\}.
\label{eq:rh12}
\end{eqnarray}
\end{widetext}
At finite bias and zero temperature
direct integration provides the real ($\Re$) and imaginary ($\Im$)
parts of $\rho_{1,2}$ ($\phi\neq 2\pi n$)
\bea
\Re \rho_{1,2} &=& \frac{1}{4\pi}\Big\lbrack \tan
^{-1}\Big(\frac{\mu_L-\epsilon}{\gamma_{+}}\Big)
-\tan ^{-1}\Big(\frac{\mu_L-\epsilon}{\gamma_{-}}\Big) \nonumber \\
&+& \tan
^{-1}\Big(\frac{\mu_R-\epsilon}{\gamma_{+}}\Big)-\tan
^{-1}\Big(\frac{\mu_R-\epsilon}{\gamma_{-}}\Big)
\Big\rbrack,
\eea
and
\bea
\Im \rho_{1,2} &=& \frac{1}{4\pi}\sin(\phi/2)\Big\lbrack \tan
^{-1}\Big(\frac{\mu_L-\epsilon}{\gamma_{+}}\Big)
+\tan^{-1}\Big(\frac{\mu_L-\epsilon}{\gamma_{-}}\Big) \nonumber \\
&-& \tan
^{-1}\Big(\frac{\mu_R-\epsilon}{\gamma_{+}}\Big)-\tan
^{-1}\Big(\frac{\mu_R-\epsilon}{\gamma_{-}}\Big)
\Big\rbrack.
\eea
As before, we define
$\gamma_{\pm}=\frac{\gamma}{2}(1\pm\cos\frac{\phi}{2})$. We now
readily confirm that at the symmetric point the real part vanishes
and ``phase localization" takes place \cite{Tu}. In particular, in
the infinite bias limit we find $\Im
\rho_{1,2}=\frac{1}{2}\sin\frac{\phi}{2}$, in agreement with
previous studies \cite{ABSalil}. We also include the behavior at the special
points $\phi=2\pi n$.   Eq. (\ref{eq:rh12}) reduces then to a
simple Lorentzian form, at zero temperature,
\bea
&&\rho_{1,2}(\phi=0)
=
\frac{\gamma}{4\pi}\int_{-\infty}^{\mu_L-\epsilon}  \frac{dx}{x^2+\gamma^2}
+ \frac{\gamma}{4\pi}\int_{\epsilon-\mu_R}^{\infty}  \frac{dx}{x^2+\gamma^2}
\nonumber\\
&&= \frac{1}{4\pi} \left[ \tan^{-1}\left(\frac{\mu_L-\epsilon}{\gamma} \right)
 + \tan^{-1}\left(\frac{\mu_R-\epsilon}{\gamma} \right) \right]
+\frac{1}{4}.
\nonumber\\
\eea
The sign reverses for $\phi=\pm 2\pi$. We note that the imaginary part of the
coherence identically vanishes at zero phase while the real part is
finite, approaching the value $1/4$ at the symmetric point.

\begin{figure}
\hspace{2mm} 
{\hbox{\epsfxsize=85mm \epsffile{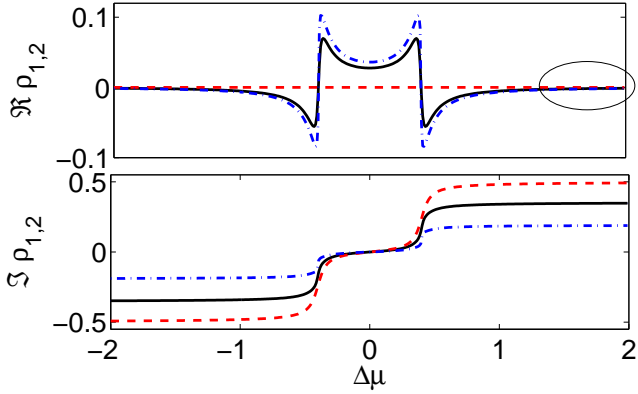}}} \caption{Real and
imaginary parts of the coherence as a function of the bias voltage.
$\phi=\pi$ (dashed line), $\phi=\pi/2$ (full line), $\phi=\pi/4$
(dashed-dotted line). Other parameters are $\epsilon=0.2$, 
$\gamma=0.05$, and $T=0$.
The oval shape marks the region of
phase localization at positive bias.  } \label{Figc1}
\end{figure}

Numerical results in the steady-state limit are displayed in Fig.
\ref{Figc1}. We find that both the real and imaginary parts of
$\rho_{1,2}$ demonstrate significant features when the dots' levels
cross the bias window, at $\Delta \mu=2\epsilon$. The value of the
real part abruptly changes sign, the imaginary part develops a step.
At large bias $\Re \rho_{1,2}$ diminishes while $\Im \rho_{1,2}$ is
finite, indicating on the development of the phase localization behavior. It can be shown that
the double-step structure of $\Im \rho_{1,2}$  (as a function of $\Delta
\mu$) disappears when the dots energies are set at the symmetric point.


\subsubsection{Current}

It is of interest to complement our study of subsystem (dots)
properties and examine the transmission coefficient and the
overall charge current in the system. The electric current, flowing
from the $L$ metal to the $R$ end, is obtained by defining the
number operator $N_L\equiv\sum_{l}a_l^{\dagger}a_l$, providing the
current $J_{L\rightarrow R}=-\frac{dN_L}{dt}=-i[H,N_L]$. This yields
\bea
J_{L\rightarrow R}=i\sum_{l,\alpha=1,2} 
\left(
 \xi_{\alpha,l}^*e^{-i\phi_{\alpha}^L}\langle a_{l}^{\dagger}a_{\alpha} \rangle
-\xi_{\alpha,l}e^{i\phi_{\alpha}^L}\langle a_{\alpha}^{\dagger}a_{l} \rangle
\right)
\eea
Expectation values are calculated in the steady-state limit.
Using the EOM formalism as explained in Sec. III, we get the standard result \cite{Wingreen}
\bea
J_{L\rightarrow R}=\frac{1}{2\pi} \int_{-\infty}^{\infty}d\omega \mathcal T_{LR}(\omega)
[f_L(\omega)-f_R(\omega)].
\label{eq:curr}
\eea
The transmission coefficient is defined as $\mathcal T_{LR}={\rm
Tr}(\Gamma^LG^+\Gamma^RG^-)$, where the trace is performed over the
states of the subsystem (dots). In the present model, at zero
temperature, we obtain
\begin{eqnarray}
&&J_{L\rightarrow R}=\frac{1}{2\pi}\int_{\mu_L}^{\mu_R}
d\omega\frac{\gamma^2(\omega-\epsilon)^2\cos^2\frac{\phi}{2}}
{\left[(\omega-\epsilon)^2-\omega_0^2\right]^2+\gamma^2(\omega-\epsilon)^2} \nonumber \\
&&=\frac{\cos\frac{\phi}{2}}{2\pi}\Big\lbrack\gamma_{+}\Big\lbrace\tan ^{-1}\Big(\frac{\mu_L-\epsilon}{\gamma_{+}}\Big)-
\tan ^{-1}\Big(\frac{\mu_R-\epsilon}{\gamma_{+}}\Big)\Big\rbrace \nonumber \\
&&-\gamma_{-}\Big\lbrace\tan^{-1}\Big(\frac{\mu_L-\epsilon}{\gamma_{-}}\Big)-\tan ^{-1}\Big(\frac{\mu_R-\epsilon}{\gamma_{-}}\Big)\Big\rbrace  \Big\rbrack,
\end{eqnarray}
which agrees with known results \cite{GefenPRB}. 
Using the NEGF formalism, we could similarly investigate the shot noise
in the double-dot AB  interferometer \cite{shot}.

\begin{figure}
\hspace{2mm} 
{\hbox{\epsfxsize=60mm \epsffile{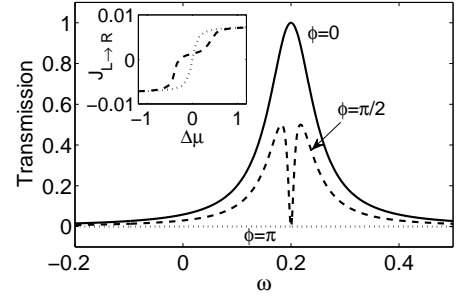}}}
\caption{Transmission coefficient
as a function of energy for $\epsilon=0.2$, $\gamma$=0.05,
$\phi=\pi$ (dotted line), $\phi=\pi/2$ (dashed line) and $\phi=0$
(full line).
The inset presents the charge current for $\phi=\pi/2$
and $\epsilon=0.2$ (dashed line), $\epsilon=0$ (dotted line). } \label{FigT}
\end{figure}

The transmission function is plotted in Fig.
\ref{FigT} displaying destructive interference pattern for
$\phi=\pi$ and a constructive behavior for $\phi=0$. For $\phi\neq
n\pi$ the transmission nullifies exactly at the position of the
resonant level \cite{Latha}. The inset presents the current-voltage
characteristics for $\phi=\pi/2$ away from the symmetric point
(dashed line), and at the symmetric point (dotted line). We note
that the double-step structure disappears at the latter case. It can
be shown that the double step structure of $\Im \rho_{1,2}$ (see
Fig. \ref{Figc1}) similarly diminishes at the symmetric point.

\section{Transient behavior}

It is of interest to investigate the development of the phase
dependency of the occupancy, and the occupancy difference $\delta
n$, before steady-state sets. Similarly, the dynamics of coherences
is nontrivial even without electron-electron interaction effects
\cite{ABSalil}. We complement the NEGF steady-state expressions of
Sec. III with numerical calculations of the transient behavior using
an exact numerical  tool that is based on the fermionic trace formula
\cite{israel}
\bea {\rm Tr}\left[e^{M_1}
e^{M_2}...e^{M_p}\right]=\det\left[1+e^{m_1}e^{m_2}...e^{m_p}\right].
\label{eq:trace} \eea
Here $m_p$ is a single-particle operator corresponding to a
quadratic operator $M_p=\sum_{i,j}(m_p)_{i,j}a_i^{\dagger}a_j$.
$a_{i}^{\dagger}$ ($a_j$) are fermionic creation (annihilation)
operators. The trace is performed over all electronic degrees of
freedom. Our objective is the dynamics of a quadratic operator
$B\equiv a_j^{\dagger}a_{k}$, $j,k=1,2$,
\bea \langle B(t) \rangle &=& {\rm Tr}\left[\rho(t_0)e^{iHt}Be^{-iHt}\right]
\nonumber\\
&=&
{\rm lim}_{\lambda \rightarrow 0}\frac{\partial}{\partial \lambda}
{\rm Tr} \left[\rho_L\rho_R\rho_S e^{iHt} e^{\lambda B}
e^{-iHt}\right]. \label{eq:exact} \eea
We introduce the $\lambda$ parameter, taken to vanish at the end of
the calculation. The initial condition is factorized,
$\rho(t_0)=\rho_S\otimes\rho_L\otimes \rho_R$,
and these density operators follow an
exponential form, $e^{M}$, with $M$ a quadratic operator. The
application of the trace formula leads to
\bea \langle e^{\lambda B(t)}
\rangle
&=&\det\Big\{[I_L-f_L]\otimes[I_R-f_R]\otimes[I_S-f_S]
\nonumber\\
&+&
 e^{iht}e^{\lambda
 b}e^{-iht} f_L\otimes f_R \otimes f_S\Big\}
\label{eq:trace2}.
\eea
with $b$ and $h$ as the single-body matrices of the $B$ and $H$
operators, respectively. The matrices $I_{\nu}$ and $I_S$ are the
identity matrices for the $\nu=L,R$ space and for the subsystem (dots). The
functions $f_L$ and $f_R$ are the band electrons occupancy
$f_{\nu}(\epsilon)=[e^{\beta(\epsilon-\mu_{\nu})}+1]^{-1}$. Here
they are written in matrix form and in the energy representation.
$f_S$ represents the initial occupation for the dots, assumed empty, again
written in a matrix form. When working with finite-size
reservoirs, Eq. (\ref{eq:trace2}) can be readily simulated
numerically-exactly.

\begin{figure}
\hspace{2mm} 
{\hbox{\epsfxsize=75mm \epsffile{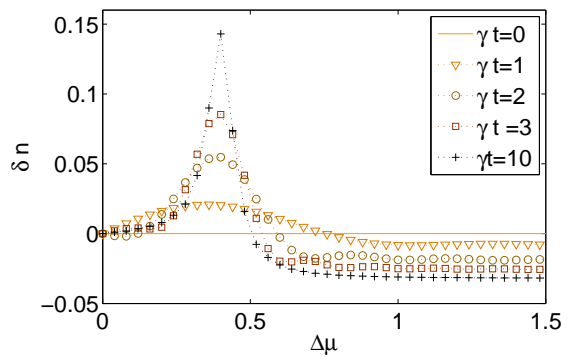}}} \caption{Time
evolution of the occupation difference, $\gamma=0.05$,
$\epsilon=0.2$, $T=5 \times 10^{-3}$, $\phi=\pi/2$.} \label{Figtime1}
\end{figure}

Fig. \ref{Figtime1} displays the evolution of the occupation
difference, presented as a function of $\Delta \mu$. In this
simulation we used finite bands with a sharp cutoff, $D=\pm 1$. At
short time $\delta n$ shows weak sensitivity to the actual bias.
Only after a certain time, $\gamma t\sim 2$, the peak around the
edge at $\Delta \mu=2\epsilon$ clearly develops. Note that 
since the
band is not very broad, edge effects are reflected at large biases
as nonzero occupation difference,
in contrast to the broad-bandwidth long-time behavior of
Fig. \ref{Figp1}.

\begin{figure}
\hspace{2mm} 
{\hbox{\epsfxsize=75mm \epsffile{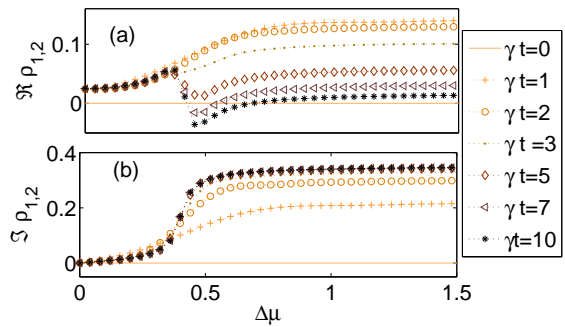}}} \caption{Time
evolution of the real (a) and imaginary (b) parts of the coherence
$\gamma=0.05$, 
$\epsilon=0.2$, $T=5 \times 10^{-3}$,
$\phi=\pi/2$. }
\label{Figtime2}
\end{figure}


The transient behavior of the coherences, $\Re \rho_{1,2}$ and $\Im
\rho_{1,2}$, is included in Fig. \ref{Figtime2}; the corresponding
steady-state value are presented in Fig. \ref{Figc1}. We can follow
the temporal features of the phase localization effect, i.e., the
disappearance of the real part of the coherence at the symmetric
point or at large bias, when $\phi\neq 2\pi n$.
Using $\phi=\pi/2$ we note that while at short to
intermediate time ($\gamma t<2$) significant coherence builds up,
the real part of the coherence eventually survives only at small
biases.
Regarding timescales, we find
that while $\Im \rho_{1,2}$ reaches the steady-state values at short
time, $\gamma t\sim2$, $\Re \rho_{1,2}$ approaches its stationary
limit only at longer times, for $\gamma t\sim 10$. Similar results were obtained in Ref. \cite{ABSalil}.

\section{Dephasing Probe: Steady state characteristics}

We have discussed so far pure coherent evolution effects in 
double-dot AB interferometers. It is important to examine at this point
the role of elastic dephasing effects on this evolution as was done 
experimentally
\cite{HeilbD1,HeilbD2} and theoretically \cite{UedaD,Zhu,KuboD} in related systems. 
Here, we are essentially focused on the effect of dephasing on the modulation of occupation
with magnetic phase. Recently, interference and decoherence
processes were studied not only in quantum dot structures
\cite{Buttiker2,Buttiker3}, but in molecular loops as well
\cite{Dhurba,MishaDhurba,Thoss,ThossE}.

Phase-breaking processes arise due to the interaction of electrons
with other degrees of freedom, e.g., with electrons, phonons and
defects. We generalize here the discussion of Sec. III, and
incorporate dephasing processes into our system phenomenologically,
by using the well established method of B\"uttiker dephasing probe
\cite{Buttiker1}. In this technique, elastic dephasing processes on
the dots are emulated by including a third terminal, $P$,
enforcing the requirement that 
the charge current towards the probe terminal, at a given electron
energy, should vanish. Thus, electrons travel to the probe and
return to the system with a different phase, while both electron
number and electron energy are conserved. This condition sets an
electron distribution within the probe. As we show below, away from
the symmetric point this distribution effectively depends on the
magnetic phase. Other phenomenological tools to incorporate
dephasing processes in mesoscopic devices are based on the
introduction of random-phase fluctuations into the scattering matrix
\cite{Pala}, or on the inclusion of damping terms into the
off-diagonal elements of the density matrix within quantum master
equations (Lindblad or Redfield) formalisms \cite{Qopen}.

Using B\"uttiker probe method, we augment the Hamiltonian
(\ref{eq:H}) with a probe, adding to the system a
noninteracting electron reservoir $P$,
\bea
H_D=H+\sum_{p\in P}\omega_pa_p^{\dagger}a_p
+ \sum_{p\in P} \lambda_{p}a_1^{\dagger} a_p + h.c.
\eea
The parameter $\lambda$ denotes the coupling strength of dot '1' to
the $P$ terminal, taken as a real number. Note that we only allow
here for local dephasing on dot '1'. One could similarly consider
models where both dots are susceptible to dephasing effects,
possibly from different sources. Following the equations-of-motion
approach as detailed in Sec. III, we arrive at the steady-state
expression for the reduced density matrix
\bea
\langle a_{\alpha}^{\dagger}a_{\beta}\rangle 
&=&\frac{1}{2\pi}\sum_{\nu=L,R,P}\int_{-\infty}^{\infty}
\left(G^{+}\Gamma^{\nu}G^{-}\right)_{\alpha,\beta}f_{\nu}(\omega)
d\omega.
\nonumber\\
\label{eq:rhoP}
\eea
The probe hybridization matrix is given by
\begin{equation} \Gamma^{P}=\gamma_P\left[ \begin{array}{cc}
              1 & 0\\
              0 & 0\\
\end{array}\right],
\end{equation}
and the dot's Green's function is written by generalizing the matrix
(\ref{eq:G1}), to include the probe self energy,
\bea G^{+}=\left[  \begin{array}{cc}
 \omega-\epsilon_{1}+\frac{i(\gamma_L+\gamma_R+\gamma_P)}{2} & \frac{i\gamma_L}{2}e^{i\phi/2}
+\frac{i\gamma_R}{2}e^{-i\phi/2}\\
\frac{i\gamma_L}{2}e^{-i\phi/2}+\frac{i\gamma_R}{2}e^{i\phi/2}
&  \omega-\epsilon_{2}+\frac{i(\gamma_L+\gamma_R)}{2}\\
 \end{array}\right]^{-1}.
\nonumber
\eea
This matrix is written here in a general form, to allow one to
distinguish between the two dots and the different dots-metals
hybridization terms. The dot-probe hybridization is defined as
$\gamma_P=2\pi\sum_p|\lambda_p|^2\delta(\omega-\omega_0)$, in
analogy with Eq. (\ref{eq:Gamma}). In our calculations below we
assume energy degenerate dots and symmetric couplings,
$\epsilon=\epsilon_1=\epsilon_2$, $\gamma_L=\gamma_R=\gamma/2$.

We now derive the probe distribution by demanding that the energy
resolved charge current to the $P$ terminal vanishes. The total
current to $P$ is given by the sum of the currents from the $L$ and
$R$ terminals,  generalizing Eq. (\ref{eq:curr}),
\bea J_P&=& J_{L\rightarrow P} +J_{R\rightarrow P}
\nonumber\\
&=&
\frac{1}{2\pi}\int_{-\infty}^{\infty}
\mathcal T_{LP}(\omega)
\left[f_L(\omega)-f_P(\omega)\right]
d\omega
\nonumber\\
&+&  \frac{1}{2\pi}\int_{-\infty}^{\infty}\mathcal T_{RP}(\omega)
\left[f_R(\omega)-f_P(\omega)\right]d\omega \eea
with the transmission coefficient $\mathcal T_{\nu\tilde
\nu}(\omega)=\rm Tr[\Gamma^{\nu}G^+\Gamma^{\tilde \nu}G^- ]$. By
requiring the integrand to vanish, we arrive at the probe
distribution
\bea f_P(\omega)=\frac{\mathcal T_{LP}(\omega)f_L(\omega) + \mathcal
T_{RP}(\omega)f_R(\omega)} {\mathcal T_{LP}(\omega) + \mathcal
T_{RP}(\omega)}. \eea
Direct evaluation of these transmission coefficients provide the electron
distribution in the probe,
\bea f_P(\omega)&=& \frac{f_L(\omega)+f_R(\omega)}{2}
\nonumber\\
&+&
\frac{\gamma(\omega-\epsilon)\sin \frac{\phi}{2}  \cos \frac{\phi}{2}  }
{2[(\omega-\epsilon)^2 + \omega_0^2]} \left[f_L(\omega)-f_R(\omega)\right].
\label{eq:fp}
\eea
As before, $\omega_0=\frac{\gamma}{2} \sin {\frac{\phi}{2}}$. This
expression indicates that the magnetic flux plays a role in setting
the distribution within the probe, (such that it only dephases the
system and does not deplete electrons or allow energy
reorganization). This dependency disappears when the dots energies
are set at the symmetric point, since the contribution of the second
term in Eq. (\ref{eq:fp}) diminishes in  the integrals of Eq. (\ref{eq:D1}), from
symmetry considerations. We now write integral expressions for the
dots occupations using Eq. (\ref{eq:rhoP}),
%
\begin{widetext}
\bea
\rho_{1,1}&=&
\frac{\gamma}{4\pi}
\int_{-\infty}^{\infty} \frac{d\omega}{\Delta(\omega)} \Bigg \{
\left[ (\omega-\epsilon)^2 +\omega_0^2 -2\omega_0(\omega-\epsilon)\cos\frac{\phi}{2}\right]f_L(\omega)
+
\left[ (\omega-\epsilon)^2 +\omega_0^2 +2\omega_0(\omega-\epsilon)\cos\frac{\phi}{2}\right]f_R(\omega)
\Bigg \}
\nonumber\\
&+&
\frac{\gamma_P}{2\pi}\int_{-\infty}^{\infty}\frac{d\omega}{\Delta(\omega)}
\left[(\omega-\epsilon)^2 +\frac{\gamma^2}{4} \right]f_P(\omega)
\nonumber\\
\rho_{2,2}&=&
\frac{\gamma}{4\pi}
\int_{-\infty}^{\infty} \frac{d\omega}{\Delta(\omega)} \Bigg \{
\left[ (\omega-\epsilon)^2 +\omega_0^2 +2\omega_0(\omega-\epsilon)\cos\frac{\phi}{2}
+\omega_0\gamma_P\sin\frac{\phi}{2}+\frac{\gamma_P^2}{4}
\right]f_L(\omega)
\nonumber\\
&+&
\left[ (\omega-\epsilon)^2 +\omega_0^2 -2\omega_0(\omega-\epsilon)\cos\frac{\phi}{2}
+\omega_0\gamma_P\sin \frac{\phi}{2} +\frac{\gamma_P^2}{4}
\right]f_R(\omega)
\Bigg \}
\nonumber\\
&+&
\frac{\gamma^2\gamma_P}{8\pi}\cos^2\frac{\phi}{2}\int_{-\infty}^{\infty}\frac{d\omega}{\Delta(\omega)}
f_P(\omega), \label{eq:D1} \eea
\end{widetext}
with
\bea
\Delta(\omega)=
\Big|(\omega-\epsilon)^2-\omega_0^2 -\frac{\gamma\gamma_P}{4} +
 i\left(\gamma+\frac{\gamma_P}{2}\right)(\omega-\epsilon)\Big|^2.
\nonumber
\eea
In the absence of dephasing these expressions reduce to Eqs. (\ref{eq:p1E}) and (\ref{eq:p2E}).
In the opposite limit, at very large dephasing, $\gamma_P\gg\gamma$, $\gamma_P> \Delta\mu$,
we note that $\rho_{2,2}$ is dominated by $\gamma_P^2\gamma$ terms that are flux independent, while
 $\rho_{1,1}$  is dominated by its last term, $\propto \gamma_P f_P$, 
which is flux dependent away from the symmetric point, resulting in
$\rho_{1,1}\propto \sin(\phi)$. Thus, quite counter-intuitively,
we find that the level that is directly susceptible to local dephasing
demonstrates flux dependency of occupation at strong dephasing, while the level that {\it indirectly} 
suffers dephasing effects more feasibly looses its coherent oscillations.

\begin{figure}
\hspace{2mm} 
{\hbox{\epsfxsize=80mm \epsffile{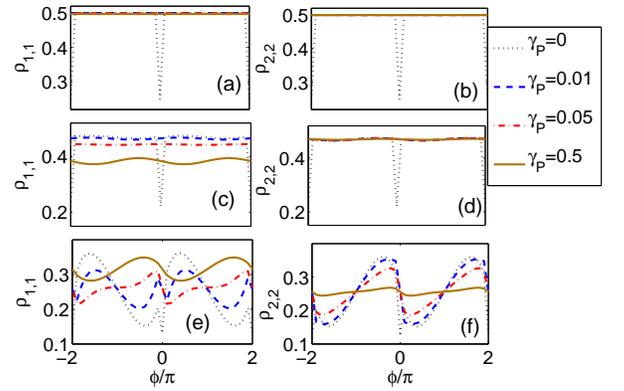}}} \caption{
The role of dephasing on the
dots occupations - magnetic phase dependency,
(a)-(b)  $\epsilon=0$,
(c)-(d) $\epsilon=0.2$, 
(e)-(f) $\epsilon=0.3$,
where $\gamma_P$=0 (dotted line), $\gamma_P=0.01$ (dashed line),
$\gamma_P=0.05$ (dashed-dotted line) and $\gamma_P=0.5$ (full line).
Other parameters are $\gamma=0.05$, $\Delta\mu=0.6$, $T=0$.}
\label{Figdeph1}
\end{figure}

Using numerical integration, 
dots occupations and their oscillation with phase are presented in
Fig. \ref{Figdeph1}. We observe the following trends upon increasing
dephasing strength: At the symmetric point, (a)-(b), the abrupt jump
at zero magnetic phase immediately disappears with the application of
finite dephasing. When the dot energies are placed away from the
symmetric point, yet they buried within the bias window, (c)-(d),
the abrupt jump at zero magnetic phase again disappears, though the
oscillations of occupation with phase prevail till large dephasing,
$\gamma_P\sim\Delta \mu$. More significantly, when the dots energies
are tuned at the edge of the bias window, (e)-(f), we find that dot
'1' (which is directly dephased) develops new type of oscillation
with phase. Only at very large dephasing, $\gamma_P\gg \Delta \mu$,
these oscillations are overly suppressed.
%
Thus, away from the symmetric point not only
features of coherent dynamics survive even at significant dephasing strength,
new type of coherent oscillations may develop as a result of the
application of elastic scattering effects on the dots.
It is interesting to reproduce this behavior while
modeling elastic dephasing effects using other techniques \cite{UedaD,Zhu,KuboD,Pala,Qopen}.

\section{Conclusions}

In this paper, we have addressed the issue of magnetic field control
on electronic occupation and coherence in double-dot AB
interferometers. The system under investigation included energy
degenerate dots with symmetric dot-metals hybridization strengths.
However, by voltage gating the dots energies away from the so-called
symmetric point at which $\epsilon=(\mu_L+\mu_R)/2$ we have resolved
four nontrivial effects that can allow for significant
controllability over dots occupations and their coherence: (i)
Dots occupations may significantly vary with magnetic flux,
particularly when the dots levels reside close to the bias edge.
(ii) The dots acquire different occupations, though they are energy
degenerate. This behavior is maximized at the bias edge
$\epsilon\sim \mu_L$. It survives at finite temperature, as long as
$T<\gamma$. (iii) Regarding the dots coherence, we have proven that
the effect of ``phase localization" \cite{Tu} does not take place
away from the symmetric point, allowing for decoherence control in
the system. Furthermore, (iv) we have found that away from the
symmetric point the system can withstand dephasing processes,
maintaining its coherent evolution and even develop new type of
oscillations under intermediate dephasing strengths ($\gamma_P\sim
\Delta \mu$, and $\gamma_P\gg \gamma$).

Our minimal model could be applied to describe magnetic field
control in mesoscopic conducting loops and in molecular ring
structures. In the latter case it was particularly noted that
degeneracy is crucial for allowing controllability within realistic
magnetic fields \cite{Dhurba}. Our study has been limited to the
noninteracting electron model, excluding both electron-electron
interaction effects and other explicit sources for dephasing and
inelastic scattering processes. It is of interest to explore the
role of interactions on the effects revealed in this paper, as we expect it
to lift the energy degeneracy in the system, further intensifying the effects discussed here.
This behavior can be immediately
observed at the mean-field level. 
The Hartree term corrects the dots energies, e.g.
 $\epsilon_1 \rightarrow \epsilon_1+U\rho_{2,2}$ \cite{MF}. Thus, away from
the symmetric point, 
(flux generated) unequal dots occupations 
translate to effective unequal energetics for the two (identical) dots.

\acknowledgments DS acknowledges support from NSERC discovery grant.
The research of SB was supported by an Early Research Award of DS.
The authors acknowledge Savannah Garmon for useful discussions.


\renewcommand{\theequation}{A\arabic{equation}}
\setcounter{equation}{0}  
\section*{Appendix A: Derivation of Equation (\ref{eq:F1})}

In this appendix our aim is to evaluate the following integral analytically
\bea
I=\frac{\gamma}{4\pi}\int_{\mu_R}^{\mu_L}d\omega
\frac{2\omega_0(\omega-\epsilon)\cos\frac{\phi}{2}}
{\left[(\omega-\epsilon)^2-\omega_0^2\right]^2+ [\gamma(\omega-\epsilon)]^2 }.
\label{eq:AF1}
\eea
We achieve this by using the following definite integral
\bea
&& I_0=\int_d^c \frac{x}{(x^2-a^2)^2+b^2x^2}dx
\nonumber\\
&&=
\frac{\tan^{-1}\left[ \frac{2a^2-b^2-2d^2}{b\sqrt{4a^2-b^2}}\right]
-
\tan^{-1}\left[ \frac{2a^2-b^2-2c^2}{b\sqrt{4a^2-b^2}}\right]}
{b\sqrt{4a^2-b^2}}
\label{eq:A1}
\eea
where $d=(\mu_R-\epsilon)$, $c=(\mu_L-\epsilon)$, $b=\gamma$ and 
$a=\frac{\gamma}{2}\sin \frac{\phi}{2}$, leading to
$b\sqrt{4a^2-b^2}=\pm i\gamma^2 \cos \frac{\phi}{2}$ and
\bea
&&2a^2-b^2-2d^2= \gamma^2\left[\frac{1}{2}\sin^2\frac{\phi}{2} -1\right] -2(\mu_R-\epsilon)^2
\nonumber\\
&&2a^2-b^2-2c^2= \gamma^2\left[\frac{1}{2}\sin^2\frac{\phi}{2} -1\right] -2(\mu_L-\epsilon)^2
\nonumber
\eea
We now reorganize Eq. (\ref{eq:A1}) using the relations
$\tan^{-1}x +\tan^{-1}y = \tan^{-1}\left(\frac{x+y}{1-xy}\right)$
and
$\tan^{-1}z=\frac{i}{2}\left[ \ln(1-iz) - \ln(1+iz)\right]$,
to find 
%
\bea
I_0=\frac{\ln \left[\frac{F_+(\phi)}{F_-(\phi)}\right] }{2\gamma^2 \cos \frac{\phi}{2}},
\eea
where we define
\bea
&&F_{\pm}(\phi)=\frac{\gamma^4}{8}\sin^4\frac{\phi}{2}
\nonumber\\
&&-
(\mu_L-\epsilon)^2
\Big[\frac{\gamma^2}{2}\sin^2\frac{\phi}{2}-(\mu_R-\epsilon)^2-\gamma^2\left(1\pm\cos\frac{\phi}{2}\right)\Big]
\nonumber\\
&&-(\mu_R-\epsilon)^2\left[\frac{\gamma^2}{2}\sin^2\frac{\phi}{2}-
(\mu_L-\epsilon)^2-\gamma^2\left(1\mp\cos\frac{\phi}{2}\right)\right].
\nonumber\\
\label{eq:AF2}
\eea
We can also reorganize these factors as a sum of real quadratic terms,
\bea
F_{\pm}(\phi)&=& \frac{\gamma^4}{8}\sin^4 \frac{\phi}{2}+2(\mu_L-\epsilon)^2(\mu_R-\epsilon)^2
\nonumber\\
&+&\frac{\gamma^2}{2}\left(\cos\frac{\phi}{2}\pm 1\right)^2 (\mu_L-\epsilon)^2
\nonumber\\
&+&\frac{\gamma^2}{2}\left(\cos\frac{\phi}{2}\mp 1\right)^2 (\mu_R-\epsilon)^2.
\label{eq:AF3}
\eea
Attaching the missing prefactors, 
$I=\frac{\gamma}{4\pi} 2\omega_0 \cos\frac{\phi}{2}I_0$, we obtain Eq. (\ref{eq:F1})
\bea
I=\frac{\sin\frac{\phi}{2}}{8\pi} \ln \left[\frac{F_+(\phi)}{F_-(\phi)}\right].
\eea


\end{document}